\documentclass{elsart}

\usepackage{graphics,natbib,graphicx,psfig,amssymb}
\journal{New Astronomy}
\input psfig.sty

\begin{document}

\begin{frontmatter}

\title{Implication of Existence of Hybrid stars and Theoretical Expectation of Submillisecond Pulsars}

\author[chinese]{Xiaoping Zheng\corauthref{cor}},
\corauth[cor]{corresponding author.} \ead{zhxp@phy.ccnu.edu.cn}
\author[chinese]{Nana Pan},
\author[chinese]{Shuhua Yang},
\author[chinese]{ Xuewen Liu},
\author[chinese]{Miao Kang}
\author[chine]{and  Jiarong Li}
\address[chinese]{Institute of Astrophysics, Huazhong Normal University,
              Wuhan 430079, P. R. China}
\address[chine]{Institute of Particle Physics, Huazhong Normal University,
              Wuhan 430079, P. R. China}

\begin{abstract}
We derive the bulk viscous damping timescale of hybrid stars,
neutron stars with quark matter core. The r-mode instability
windows of the stars show that the theoretical results are
consistent with the rapid rotation pulsar data, which may give an
indication for the existence of quark matter  in the interior of
neutron stars. Hybrid stars instead of neutron or strange stars
may  lead to submillisecond pulsars.
\end{abstract}

\begin{keyword}
dense matter --- gravitation ---stars: neutron--- stars:rotation
--- stars:oscillations
\PACS 97.60.Jd \sep 12.38.Mh \sep 97.60.Gb
\end{keyword}

\end{frontmatter}

Probing the composition of matter in the interior of compact stars
is of interest for understanding of pulsars. The available
theories have not yet given a unique solution to this problem.
Since the deconfinement transition from hadron matter into quark
matter is possible at high density, one logically proposes that
some pulsars could be strange stars. But how to distinguish
strange stars from neutron stars is a difficult and an urgent
issue. First, one suggested that bulk properties of neutron stars
such as radius and moment of inertia, may be marked off due to the
difference between self-bound and gravity-bound matter.
Unfortunately, the bulk properties of these two kinds of stars are
similar in the observed mass range $1<M/M_{\odot}<2$\citep{hae86}.
Another way is to look at the possible difference of their highest
rotation frequency, since the discovery by Andersson, Friedman and
Morsink of r-mode instabilities in the rotating compact stars  put
rather severe limits on the highest rotation frequency of
pulsars\citep{and98, fri98}. A pioneer investigation found that
the limiting rotation of strange stars is in millisecond period
range but for normal neutron stars in the period range of 15$\sim$
20ms\citep{mad98, mad00}. However,  a strange star evolving to the
limiting frequency corresponds to temperature in $10^6--10^7$K
range instead of the inferred  core temperature of millisecond
pulsars of a few times $10^8$K. Obviously, this is a serious
drawback if we attributed the uncorrespondence to the
under-representing statistics due to a low number of objects as
suggested by Madsen\citep{mad00}.
 Although some works also argued that neutron stars with
solid crusts would be  the candidates to explain
  the rapidly rotating accreting pulsar's data   through thermal runaway recycle of
  r-mode. Due to the presence of viscous boundary layer
  damping\citep{bil00,and00}, the resulting viscous heating  is so
intense that it can heat the crust-core interface to the melting
temperature of the solid crust when the r-mode amplitude is larger
than some critical value, which is crudely estimated to be $\sim
10^{-3}$ by \citet{owe99} and  a more accurate value is given as
$5\times 10^{-3}$ by \citep{lin00}.

Following the above train of thought, we will here investigate the
r-mode instability window of hybrid stars, i.e.,  the neutron
stars containing quark matter core. We show that the instability
window of hybrid stars explains  the millisecond pulsar data very
well, but the strange stars and normal neutron stars would not fit
because of the weaknesses concerned  above.

In rotating relativistic stars, gravitational radiation drives the
r-mode while various dissipation mechanisms counteract the fluid
motion. In general, shear viscosity, or surface rubbing that could
also be applied to the hybrid stars, which also have a solid crust
similar to the normal neutron stars,
 suppresses the mode at low temperatures but bulk viscosity
dominates at high temperatures. The critical rotation frequency
for a given stellar model as a function of temperature follows
\begin{equation}
{1\over\tau_{gr}}+{1\over\tau_{sv}}+{1\over\tau_{bv}}+{1\over\tau_{sr}}=0,
\end{equation}
where $\tau_{gr}<0$ is the characteristic timescale for energy
loss due to gravity wave emission, $\tau_{sv}$ and $\tau_{bv}$ are
the damping times due to shear and bulk viscosities, and
$\tau_{sr}$ is that for surface rubbing due to the presence of the
viscous boundary layer.

A polytropic equation of state with a low index $n$ is a good
approximation for compact stars, as discussed in many
papers\citep{kok99,lin99}. The  timescale for $n=1$ polytrope from
gravity wave emission can be written as
\begin{equation}
\tau_{\rm gr}=-47{\rm s}\times \left(M\over
1.4M_\odot\right)^{-1}\left(R\over 10{\rm
km}\right)^{-4}\left(1000\over \nu\right)^6,
\end{equation}
where $M, R, \nu$  represent the mass, radius and the rotation
frequency of the star respectively. In general, the shear
viscosity and surface rubbing dominate at low temperature, but in
fact we can neglect the shear viscosity because the dissipation
associated with a viscous boundary layer would greatly exceed that
of the standard shear viscosity by typical $10^5$ \citep{bil00,
and00}. The timescale due to surface rubbing is given
by\citep{and00}
\begin{equation}
\tau_{sr}=200{\rm s}\times\left(M\over
1.4M_\odot\right)\left(R\over 10{\rm km}\right)^{-2}\left(T\over
10^8{\rm K}\right)\left(1000\over \nu\right)^{1/2}.
\end{equation}

We now  discuss the bulk viscosity coefficients of strange quark
matter, hadron matter and their mixed phase matter to estimate the
bulk viscous damping timescale in hybrid stars. The bulk
viscosities are given approximately by
\begin{equation}
\zeta_{QP}=3.2\times 10^{21}{\rho T_9^2\over\kappa^2\Omega^2}\left
({m_s\over 100{\rm Mev}}\right )^4
\end{equation}
for strange quark matter\citep{mad92}, and
\begin{equation}
\zeta_{HP}=6\times 10^{-5}{\rho^{2} T_9^6\over\kappa^2\Omega^2}
\end{equation}
for hadron matter\citep{ips91}, where the constant  $ \kappa=2/3$,
the angular velocity $\Omega=2\pi\nu$ and $\rho$ is the density,
$m_s$ is strange quark mass. To construct hybrid stars, we
consider the deconfinement transition occurring at high density
inside the compact stars and use BPS\citep{bps71} and
GPS\citep{gho95} equations of state for hadrons and MIT model for
quarks. As in most cases, the variation of volume per unit mass
and the chemical potential imbalance due to the oscillation in the
star is so small that it is the liner part that contributes most
to the viscosity, so we have used the relaxation time
approximation method\citep{hb02,hb05,hb055} to compute the  bulk
viscosity of the mixed phase matter,
\begin{equation}
 \zeta=\frac{P(\gamma_{\infty}-\gamma_{0})\tau}{1+(\omega\tau)^{2}}.
  \end{equation}
 Here $P$
is the pressure of the mixed phase, $\gamma_{\infty}$ and
$\gamma_{0}$ are the "infinite" and "zero" frequency adiabatic
index, $\omega$ is the angular velocity of the perturbation and
$\tau$ is the relaxation time associated to the weak processes. As
an example, we numerically calculate the bulk viscosity
coefficient of the mixed phase matter which is showed in figure 1
for  bag constant $B^{1/4}=165 $ Mev. From numerical estimate, we
easily find the fact that
$\zeta_{QP}\approx\zeta_{MP}\gg\zeta_{HP}$.
 \begin{figure}
 \includegraphics[width=0.9\textwidth]{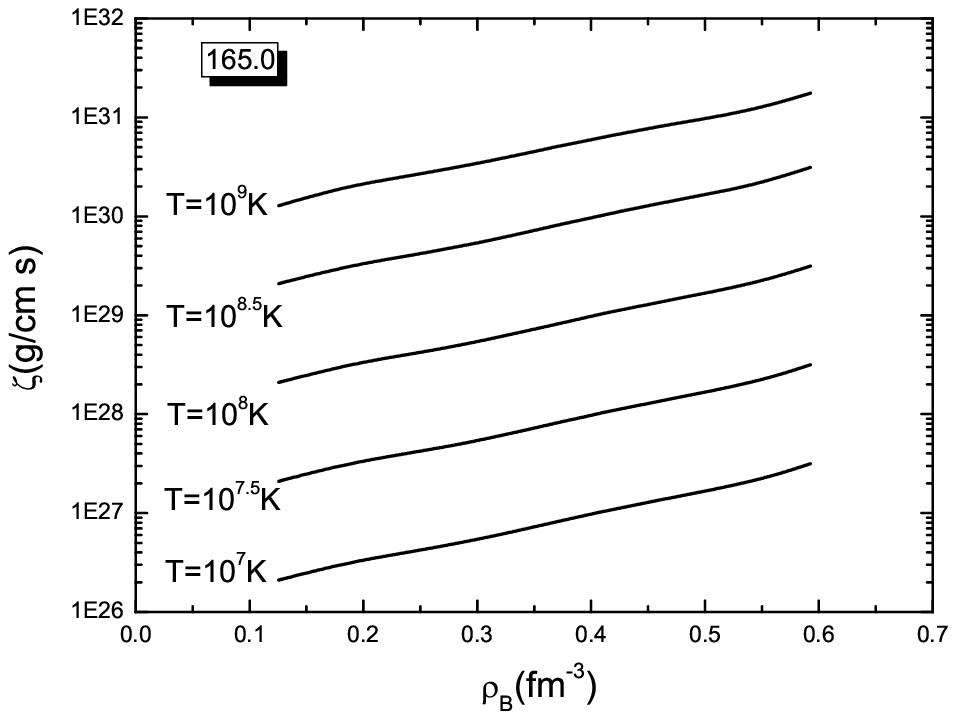}
 \caption[]{The bulk viscosity of the mixed phase matter in
the hybrid star for $m_{s}=150Mev $ and $B^{1/4}=165 $Mev in terms
of GPS model\citep{gho95}. \label{fig2}}
 \end{figure}

 Now, we can use the bulk viscosities listed above to evaluate the bulk viscous damping times for the hybrid stars.
  For the given energy of modes $E$ and its variation rate $ \dot{E}\equiv {\rm d}E/{\rm d}t$, the timescale is defined as
  \begin{equation}
  {1\over\tau}=-{\dot{E}\over 2E}.
  \end{equation}
  The r-mode energy in a rotating relativistic star with radius $R$ has been expressed as\citep{lin99}
  \begin{equation}
  E={\alpha^2\pi\over 2m}(m+1)^3(2m+1)!R^4\Omega^2\int_0^R\rho(r)\left ({r\over R}\right )^{2m+2}{\rm d}r+O(\Omega^4),
  \end{equation}
  where $\alpha$ is the normalization parameter,  $m$ is integer and $\rho(r)$ is the density function\citep{and01}.
  The energy dissipation rate due to bulk viscosity can be estimated with the
  following formula
  \begin{equation}
  \dot{E}_{bv}=-\int\zeta\delta\sigma\delta\sigma^*{\rm d}^3x,
  \end{equation}
where  $\delta\sigma$ is expansion tensor of Eulerian perturbation
of velocity field, expressed approximately as\citep{lin99}
 \begin{equation}
  \delta\sigma=-
  i(\omega+m \Omega)\frac{\delta\rho}{\rho},
  \end{equation}
here $\omega$ denotes angular frequency of the r-mode. In
lowest-order approximation, we easily find $\delta\sigma\sim
R^{2}(\frac{r}{R})^{m+1}$.In a stratiform star, we can decompose
the energy $E$ into four terms, i.e., the quark core, mixed phase,
nuclei envelope and solid crust  denoted respectively by $QP, MP,
HP $ and $CR$
\begin{equation}
E=\left ({R_{QP}\over R}\right
)^{2m-2}(\tilde{E}_{QP}+\tilde{E}_{MP}+\tilde{E}_{HP}+\tilde{E}_{CR})
\end{equation}
with
\begin{equation}
\tilde{E}_{QP}={\alpha^2\pi\over
2m}(m+1)^3(2m+1)!R_{QP}^4\Omega^2\int_0^{R_{QP}}\rho_{QP}(r)\left
({r\over R_{QP}}\right )^{2m+2}{\rm d}r,
\end{equation}
\begin{equation}
\tilde{E}_{MP}={\alpha^2\pi\over
2m}(m+1)^3(2m+1)!R_{QP}^4\Omega^2\int_{R_{QP}}^{R_{MP}}\rho_{MP}(r)\left
({r\over R_{QP}}\right )^{2m+2}{\rm d}r,
\end{equation}
\begin{equation}
\tilde{E}_{HP}={\alpha^2\pi\over
2m}(m+1)^3(2m+1)!R_{QP}^4\Omega^2\int_{R_{MP}}^{R_{HP}}\rho_{HP}(r)\left
({r\over R_{QP}}\right )^{2m+2}{\rm d}r,
\end{equation}
\begin{equation}
\tilde{E}_{CR}={\alpha^2\pi\over
2m}(m+1)^3(2m+1)!R_{QP}^4\Omega^2\int_{R_{HP}}^R\rho_{CR}(r)\left
({r\over R_{QP}}\right )^{2m+2}{\rm d}r.
\end{equation}
 \begin{figure}
 \includegraphics[width=0.9\textwidth]{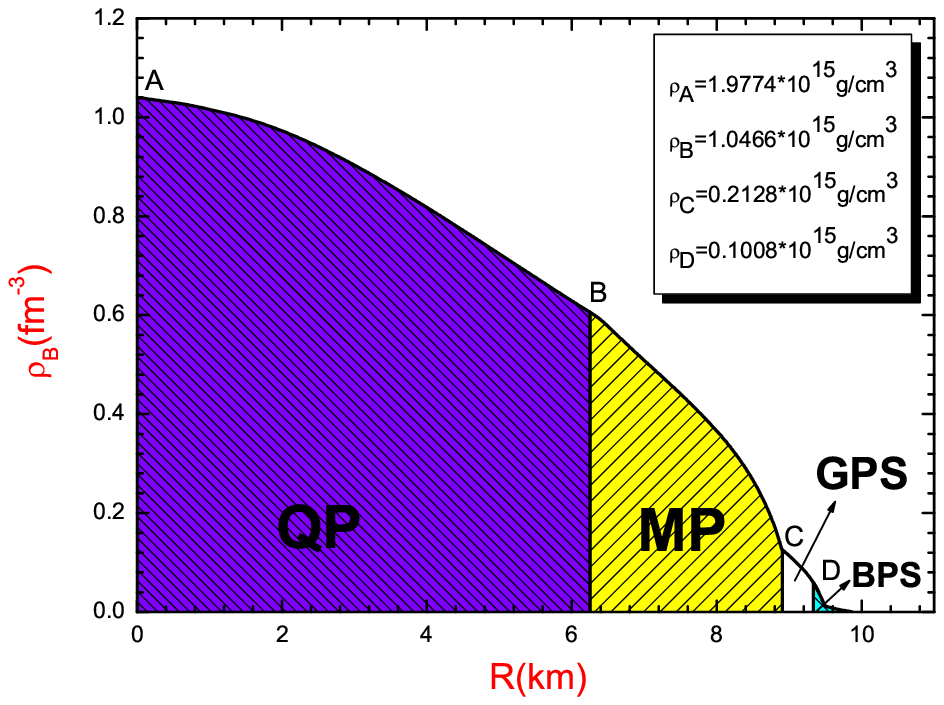}
 \caption[]{The stratiform structure of the hybrid
star with parameters $M=1.4 M_{\bigodot} $, $R=9.87km $,
$m_{s}=150Mev $ and $B^{1/4}=165Mev $ in terms of GPS
model\citep{gho95}. Here QP and MP represent the quark phase and
mixed phase  in the hybrid star respectively. \label{fig2}}
 \end{figure}
A constant density in each phase, replaced by corresponding mean
density, is a good approximation in terms of our calculations of
the stellar structure(see figure 2). Considering
$\zeta_{QP,MP}\gg\zeta_{HP,CR}$, we have, according to (8) and
(9),
\begin{equation}
\dot{E}_{bv}=(\frac{R_{QP}}{R})^{2m-2}(\dot{E}_{QP}+\dot{E}_{MP}),
\end{equation}
with
\begin{equation}
\dot{E}_{QP}=\left (dE\over dt\right )_{R_{QP}, \rho_{QP}},
\end{equation}
\begin{equation}
\dot{E}_{MP}=\left (R_{MP}\over R_{QP}\right )^{2m-2}\left
(dE\over dt\right )_{R_{MP}, \rho_{MP}}-\left (dE\over dt\right
)_{R_{QP}, \rho_{MP}}.
\end{equation}
Thus, the timescale from Eq(4) is immediately expressed as
\begin{equation}
\tau_{bv}=A\tau_{bv}^{QP}(R),
\end{equation}
where $A$ is a constant related to the stellar structure that is
showed in table 1, $\tau_{bv}^{QP}(R)$ can be given
by\citep{and01}
\begin{equation}
\tau^{QP}_{bv}=7.9{\rm s}\times\left({M\over 1.4 M_\odot}\right
)^{2}\left({R\over 10{\rm km}}\right )^{-4} \left({T\over 10^9{\rm
K}}\right )^{-2}\left ({m_s\over 100{\rm MeV}}\right )^{-4}\left
({1000\over\nu}\right )^2.
\end{equation}
Inserting (2), (3) and (18), we apply Eq(1) to evaluate the
critical frequency of the rotating star as a function of
temperature. Figure 3 shows  the instability windows of the hybrid
stars for $M=1.4M_\odot$ and $R$ ranging from 10$\sim$ 14km. As
comparison, we also give the windows of strange star and normal
star in figure 3.
\begin{table}
  \caption[]{ The Values of $A$ for
different Structures of a Hybrid Star with $M=1.4 M_{\odot}$
  }
  \label{Tab:publ-works}
  \begin{center}\begin{tabular}{clcl}
  \hline\noalign{\smallskip}
  $B^{1/4}$[MeV]
      & $R$[km] & $A$                   \\
  \hline\noalign{\smallskip}
   165.0     & 9.87  & 0.6886         \\
   175.0     & 10.41 & 0.6929 \\
   185.0     & 12.67 & 6.5488\\
  \noalign{\smallskip}\hline
  \end{tabular}\end{center}
\end{table}

 \begin{figure}
 \includegraphics[width=0.9\textwidth]{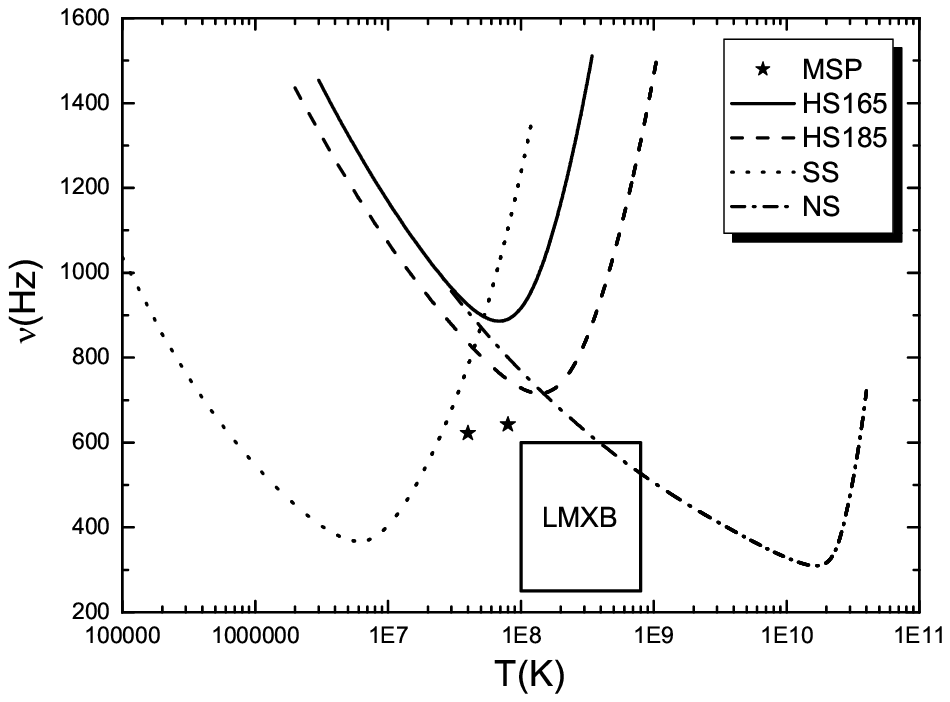}
 \caption[]{Comparison of r-mode instability
windows of hybrid stars with  several other
models\citep{and00,mad00,zh03}. HS, SS and NS denote hybrid star,
strange star  and neutron star respectively, while the star symbol
and the rectangle indicate the two observed fastest pulsars and
the Low Mass X-rays Binaries. \label{fig3}}
 \end{figure}

For hybrid stars, one notes that a pulsar will reach the frequency
of about 700 Hz( 1.43 ms) following a track coincident with the
curve marking the instability region. So, the LMXB's are well
within the region stable against r-mode instabilities, just
spinning  up the region unhindered by the instability due to
accretion.  For strange stars, there is the same mechanism as
hybrid star model but a deviation of  temperature from the pulsar
data\citep{mad00,zh03}. It is accepted that the neutron stars with
crust could also fit the pulsar data through r-mode runaway
recycle mechanism \citep{lev99,bil00,and00}. An important point is
that, the fate of the crust of neutron stars is different from
that of hybrid stars: the former is melted probably because of
r-mode dissipation due to viscous boundary layer. We can  imagine
that the existence of the higher rotation pulsars is possible if
there would exist larger quark matter core in the interior of a
compact star.  This may imply that submillisecond pulsars, if they
exist, should be hybrid stars instead of neutron stars or strange
stars.

In conclusion, the bulk viscous damping timescale for hybrid stars
is estimated for a range of parameters through which the r-mode
instability window is obtained. The rapid rotation pulsar data is
a significant indication of the existence of hybrid stars instead
of neutron stars or strange stars. Naturally, this is a possible
evidence of quark matter in the interior of the compact stars.

We would like to thank the  support by National Natural Science
Foundation of China under Grant No. 10373007, 90303007 and the
Ministry of Education of China with project No. 704035.  We are
especially grateful to Professor D.F. Hou for reading and checking
the text.

\end{document}